\documentclass[journal]{IEEEtran}
\usepackage{algorithm}
\usepackage{algorithmic}
\usepackage{multirow}
\usepackage{hyperref}
\usepackage{cite}
\usepackage{graphicx,color}
\usepackage{graphicx}
\usepackage{amsmath}

\begin{document}

\title{Dynamic Predictive Sampling Analog to Digital Converter for Sparse Signal Sensing}

\author{Xiaochen~Tang,~\IEEEmembership{Member,~IEEE,},
Mario Renteria-Pinon,  and~Wei~Tang,~\IEEEmembership{Member,~IEEE}
\thanks{Manuscript received July 7, 2022; revised XXX XX, 2022. This work was supported by the United States National Science Foundation under Grant Nos. ECCS-1652944, and ECCS-2015573.} 
\thanks{Xiaochen Tang, Mario Renteria-Pinon, and Wei Tang are with the Klipsch School of Electrical and Computer Engineering, New Mexico State University, Las Cruces, NM, USA.}
\thanks{Corresponding author: Wei Tang, e-mail: wtang@nmsu.edu}}
\maketitle
\begin{abstract}
This paper presents a dynamic predictive sampling (DPS) based analog-to-digital converter (ADC) that provides a non-uniform sampling of input analog continuous-time signals. The processing unit generates a dynamic prediction of the input signal using two prior-quantized samplings to compute digital values of an upper threshold and a lower threshold. The digital threshold values are converted to analog thresholds to form a tracking window. A comparator compares the input analog signal with the tracking window to determine if the prediction is successful. A counter records timestamps between the unsuccessful predictions, which are the selected sampling points for quantization. No quantization is performed for successfully predicted sampling points so that the data throughput and power can be saved. The proposed circuits were designed as a 10-bit ADC using 0.18 micro CMOS process sampling at 1 kHz. The results show that the proposed system can achieve a data compression factor of 6.17 and a power saving factor of 31\% compared to a Nyquist rate SAR ADC for ECG monitoring. 
\end{abstract}


\IEEEpeerreviewmaketitle

\section{Introduction}\label{Introduction}

\begin{figure}[t!]
	\begin{center}
		\includegraphics[width=3.0in]{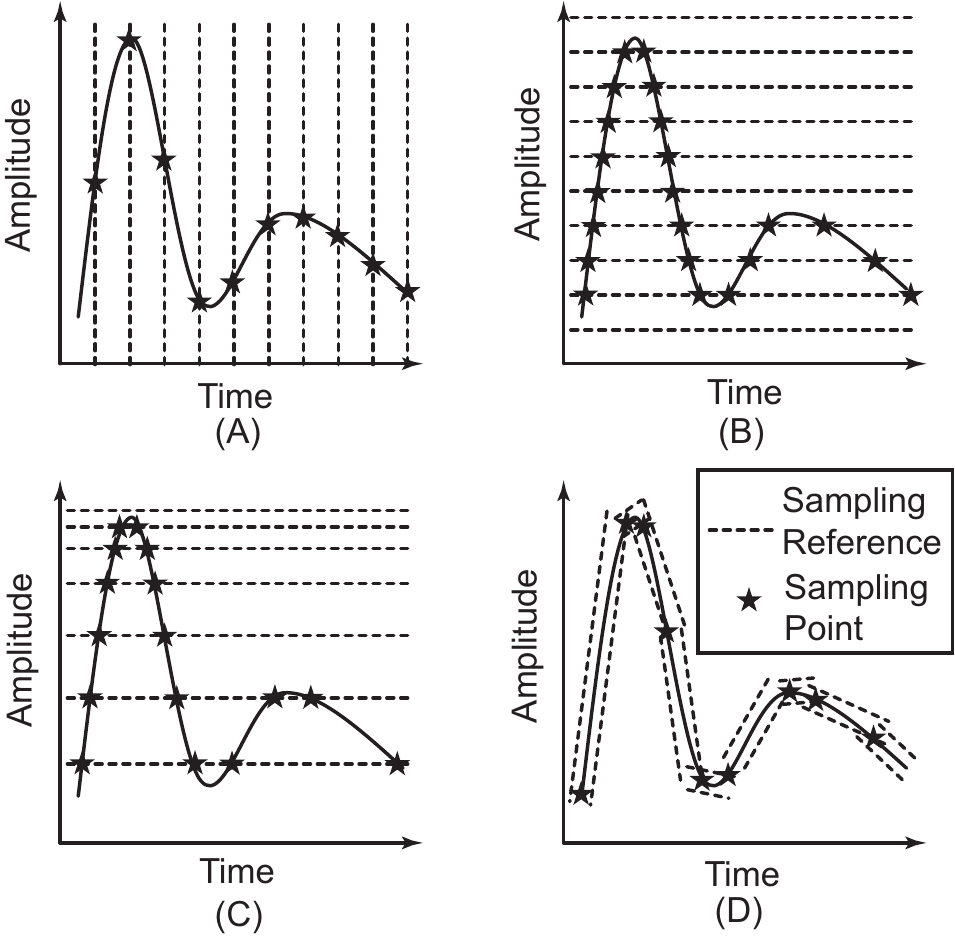}
	\end{center}
	\vspace{-0.2in}
	\caption{Comparing different sampling methods: (A) Nyquist sampling, (B) Level-crossing sampling, (C) Logarithmic level-crossing sampling, and (D) Dynamic predictive sampling (DPS). The proposed DPS selects the least number of sampling points compared to other sampling methods.}
	\label{fig:sampling}
	\vspace{-0.2in}
\end{figure}

\IEEEPARstart{L}{ow}-power sensing hardware and algorithms for data acquisition systems are critical for wearable and miniaturized devices and have been actively studied. The primary goal is to convert the input analog signal into digital data while extracting the critical information from the input signal and avoiding unnecessary data generated from the conventional Nyquist sampling. As shown in Fig. \ref{fig:sampling} (A), the conventional Nyquist sampling uses a fixed sampling clock and converts each sampling point into digital values. Such a method generates too much data for the following digital signal processing circuits and systems, which introduces high power consumption for both signal processing and communication. In particular, since many signals in Internet-of-Things (IoT) and biomedical applications are sparse in the time domain, only the active portion or the spikes in the signal are of interest. Thus, Nyquist sampling consumes too much power in sampling and quantization while generating  unnecessary data.  

To address this issue, nonuniform sampling methods are proposed. The most popular solution is the event-based level-crossing sampling \cite{ZhaoLianTCASII2022} that samples using amplitude thresholds instead of a constant clock \cite{SchellTsividisJSSC2008,TangOsmanCulurcielloTCASI2013,LiZhaoSerdijnTBCAS2013,WeltinTsividisJSSC2013,HouYousefLianTCASII2018} as shown in Fig. \ref{fig:sampling} (B). In this method, if the input analog signal’s amplitude variation is below a certain threshold, no sampling and quantization are performed. This is an efficient way to save power and sampling data when the input signal is sparse in the time domain. To further reduce the number of sampling points, a logarithmic level-crossing sampling method as shown in Fig. \ref{fig:sampling} (C) \cite{SirimasakulThanachayanontECTICON2017} is proposed with a cost of calculating the sampling reference voltage. However, since the event-based level-crossing sampling only generates a positive or negative pulse when the input signal is triggering a few prior defined voltage levels, it suffers from insertion/deletion errors of the pulse sequences \cite{HuTangMWSCAS2015}, which makes it difficult for signal reconstruction. Moreover, the event-based level-crossing sampling is susceptible to high-amplitude low-frequency baseline wandering and low-amplitude high-frequency noise. In these two cases, the first-order level-crossing sampling may also perform many unnecessary samplings and generate nonessential data. Furthermore, the first-order level-crossing sampling is not good at identifying turning points (fiducial points) of the input signal, such as the timing of a peak, since it is only sensitive to the slope but not the slope variation. 


Slope-tracking sampling methods have also been proposed to reduce the number of sampling points. For instance, \cite{HafshejaniElmiMirabbasiTCASI2020} demonstrated a signal-depending sampling method that discards sampling points if the slope variation between segments of sampling points is under a certain threshold. However, the slope calculation is processed in the analog domain including calculating divisions. Such a design may suffer from noise in the analog circuits. Moreover, if the slopes of consecutive segments between sampling points have variations small enough but accumulative to one direction, the system may unnecessarily discard important sampling points, which would introduce distortion in the reconstructed signal. Therefore, additional efforts have to be made to limit the maximum number of consecutive samples that can be dropped, which increases the computing complexity. Such problems can be avoided if the sampling point selection process is performed in the digital domain after an ADC \cite{HafshejaniTaheriNejadMirabbasiSJ2022}. However, this arrangement requires a complete ADC running before digital processing at a fixed sampling rate with full-bits quantization for every sampling point. Although it could save data throughput for communication circuits, the sampling and quantization power remains the same as a Nyquist rate sampling ADC.

In this paper, we report a novel sampling method based on dynamic prediction, which selects only the important sampling points for quantization. As shown in Fig. \ref{fig:sampling} (D), the proposed method applies two tracking threshold voltages that update their value based on the slope of prior samplings. The system performs quantization only when the input signal is crossing the thresholds. This method can achieve data and power reduction as well as feature extraction during the sampling process. The sampling algorithm can be integrated into the analog to digital conversion process in a fully digital fashion without calculating the slopes. The following sections of the paper present the detail of sampling methods, simulation results, and the power and data saving analysis compared to the conventional Nyquist sampling ADCs and other nonuniform sampling ADCs.

\vspace{-0.1in}
\section{Dynamic Predictive Sampling}\label{sec:Sampling}

The dynamic predictive sampling method uses two prior sampling points to decide if the next sampling point is selected for quantization. The proposed system records the timing values and digital amplitude values of only the selected sampling points. This is performed in real-time during the sensing process to alleviate unnecessary quantization to save data and power. The following subsections describe the basic method of dynamic predictive samplings and hardware system implementation.


\begin{figure}[t!]
	\begin{center}
		\includegraphics[width=3.0in]{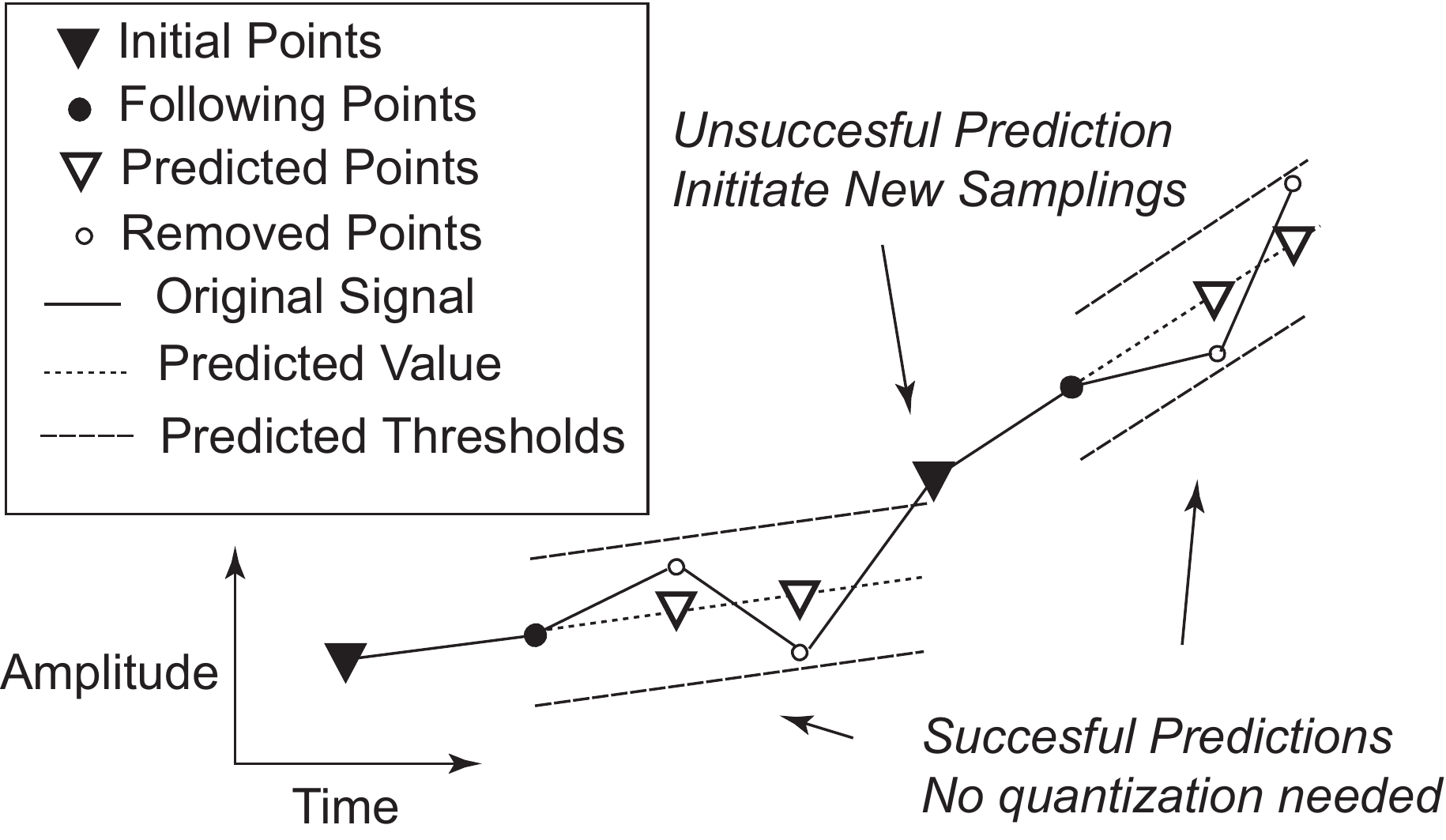}
	\end{center}
	\vspace{-0.2in}
	\caption{DPS selects sampling points and calculating predictions.}
	\label{fig:prediction}
	\vspace{-0.2in}
\end{figure}

\vspace{-0.1in}
\subsection{Prediction and Thresholds}

As shown in Fig. \ref{fig:prediction}, when the prediction process begins, the system performs analog-to-digital conversions for the first two sampling points. Then the digital prediction of the next sampling point is generated using linear extrapolation of the two prior sampling points. Specifically, the digital prediction value is calculated using twice the digital value of the last sampling point minus the digital value of the second last sampling point, as shown in Equation \ref{eq1:prediction}:



\begin{equation}
\vspace{-0.1in}
P_{D}=2\times L1_{D}-L2_{D}
\label{eq1:prediction}
\end{equation}
Here $P_{D}$ is the predicted digital value; $L1_{D}$ is the digital value of the last sampling point; $L2_{D}$ is the digital value of the second last sampling point. Since in binary data format, multiplying by two can be achieved using a left shift of the bits, Equation \ref{eq1:prediction} does not involve an actual multiplication operation. The predicted digital value is then applied to generate the upper and lower threshold digital values by adding and subtracting a pre-defined digital Delta value using Equation \ref{eq2:thresholds}

\begin{equation}
\left\{\begin{matrix}
\begin{aligned}
UT_{D}=P_{D}+\Delta_{D}\\ 
LT_{D}=P_{D}-\Delta_{D}
\end{aligned}
\end{matrix}\right.
\label{eq2:thresholds}
\end{equation}
Here $UT_{D}$ is the digital value of the upper threshold, $LT_{D}$ is the digital value of the lower threshold, and $\Delta_{D}$ is the digital Delta value.
\vspace{-0.1in}
\subsection{Sampling Decision}

The upper and lower threshold values are then converted into analog values using a digital-to-analog converter (DAC). The analog values of the thresholds are compared with the actual analog input in the next sampling. Analog comparisons are made between the analog input signal, the upper threshold value, and the lower threshold value using a comparator. The comparison results decide if the analog input signal is between the upper threshold value and the lower threshold value shown in Equation \ref{eq3:decision}.

\begin{equation}
LT_{A}<Input_{A}<UT_{A}
\label{eq3:decision}
\end{equation}
Here $LT_{A}$ is the analog value of the lower threshold, $Input_{A}$ is the analog value of the input signal, and $UT_{A}$ is the analog value of the upper threshold. $LT_{A}$ and $UT_{A}$ are generated by the DAC based on digital values $LT_{D}$ and $UT_{D}$, respectively. 

The result of Equation \ref{eq3:decision} decides whether the prediction is successful. If the input analog value is between the two thresholds, i.e., Equation \ref{eq3:decision} is valid, the prediction is correct and no quantization is performed for the input analog signal. In the next prediction, $L1_{D}$ is then replaced by the current $P_{D}$ while $L2_{D}$ is replaced by the current $L1_{D}$. Then the new $P_{D}$ is calculated using Equation \ref{eq1:prediction}. In such a case, the system doesn't record the data and no data are sent to the output. On the other hand, if the input analog value is not between the two thresholds, i.e., Equation \ref{eq3:decision} is not valid, the prediction is incorrect. This means the input analog waveform is higher than the upper threshold or lower than the lower threshold. Then quantization is performed using Successive Approximation Register (SAR) logic for the next two sampling points to generate new digital values for prediction. In such a case, the two digital values are temporarily stored and applied as $L1_{D}$ and $L2_{D}$. Then the next predicted digital value is calculated using Equation \ref{eq1:prediction}. $L1_{D}$ is sent as an output of the system. A clocked timer starts counting the timestamp between the timing of the current sample value and the next time when a prediction is incorrect. The timestamp is also a digital output of the system.

\begin{figure}[t!]
	\begin{center}
		\includegraphics[width=3.4in]{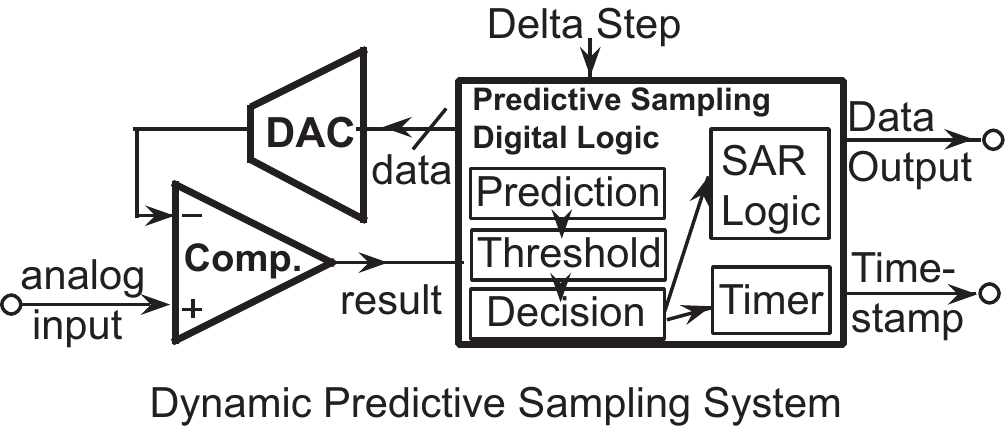}
			\vspace{-0.2in}
	\end{center}
	\caption{Block diagram of the Dynamic Predictive Sampling ADC.}
	\label{fig:system}
\end{figure}

	\vspace{-0.2in}
\subsection{Hardware Implementation}

A block diagram of the proposed dynamic predictive sampling system is shown in Fig. \ref{fig:system}, which consists of a comparator, a DAC, and a Predictive Sampling Digital Logic. The analog input signal is compared with the analog value generated from the DAC. The comparison result is sent to the digital logic for prediction and threshold calculation. The predictive sampling digital logic generates the digital data of the upper and lower threshold voltages using the predicted digital value and the Delta values. The digital data of the upper and lower thresholds are then sent to the DAC to compare with the analog input signal. If the prediction is successful, the digital logic can send the predicted digital data to the output for the debugging purpose or it can also be discarded. If the prediction is not successful, a full SAR logic is performed to obtain the digital value of the analog input. In such a case, the digital value of the analog input is sent as the Data output while a timer starts counting the clock cycles to obtain the Time-stamp output, which measures the timing difference between two unsuccessful predictions for signal reconstruction.


\vspace{-0.1in}
\section{Performance Evaluation}\label{sec:simulationResults}

\begin{figure}[!t]
\centerline{\includegraphics[width=\columnwidth]{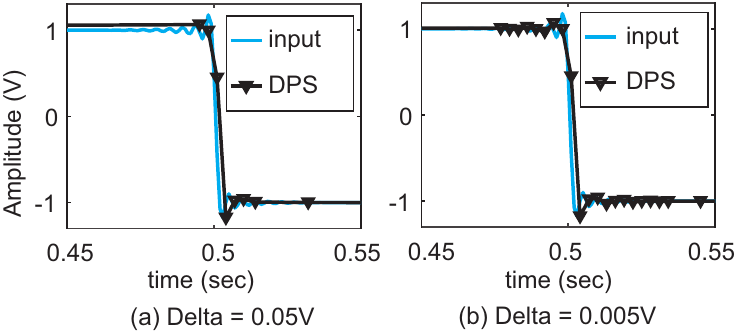}}
	\vspace{-0.1in}
\caption{Selected sampling points in low-pass-filtered square waveform with different Delta values.}
\label{fig:squrewaves}
\vspace{-0.2in}
\end{figure}

The primary goal of the dynamic predictive sampling method is to identify the key sampling points in the waveform to perform quantization. By doing so, the digital data throughput from the sensor can be reduced to alleviate processing or communication power for the following circuits in the system. In addition, the selected key sampling points also represent important features of the original waveform, which are more friendly for signal processing. The digital data from the selected sampling points should be able to reconstruct the input waveform with minimum error. Power saving is another important feature of the system especially when the input signal is sparse. In summary, the performance of the dynamic predictive sampling includes (1) data saving, (2) error introduced by reducing the number of sampling points, and (3) power saving from the analog to digital conversion steps.

Fig. \ref{fig:squrewaves} illustrates the selected sampling points using DPS with an input signal as a low-pass filtered square waveform. Such a waveform has linear portions in both the time domain and the amplitude domain, while it also contains small ripples and overshoots during the transition. As shown in Fig. \ref{fig:squrewaves}, the DPS method selects a small number of sampling points when the input signal is linear in both amplitude and time domain, which saves a lot of data compared to Nyquist sampling and level-crossing sampling. More importantly, DPS automatically selects more sampling points in the fine structure of the waveform, which preserves the key information. Both Nyquist sampling and level-crossing sampling do not have such a feature. Comparing Fig. \ref{fig:squrewaves} (a) and (b), DPS can control the trade-off between errors and data savings by using different Delta levels. The overall performance of DPS depends on specific input signals in terms of time domain sparsity, amplitude, and the selection of the Delta value. The following subsections analyze the performance trade-offs with specific signals.

\vspace{-0.2in}
\subsection{Data Savings}

\begin{figure}[!t]
	\centerline{\includegraphics[width=3.4in]{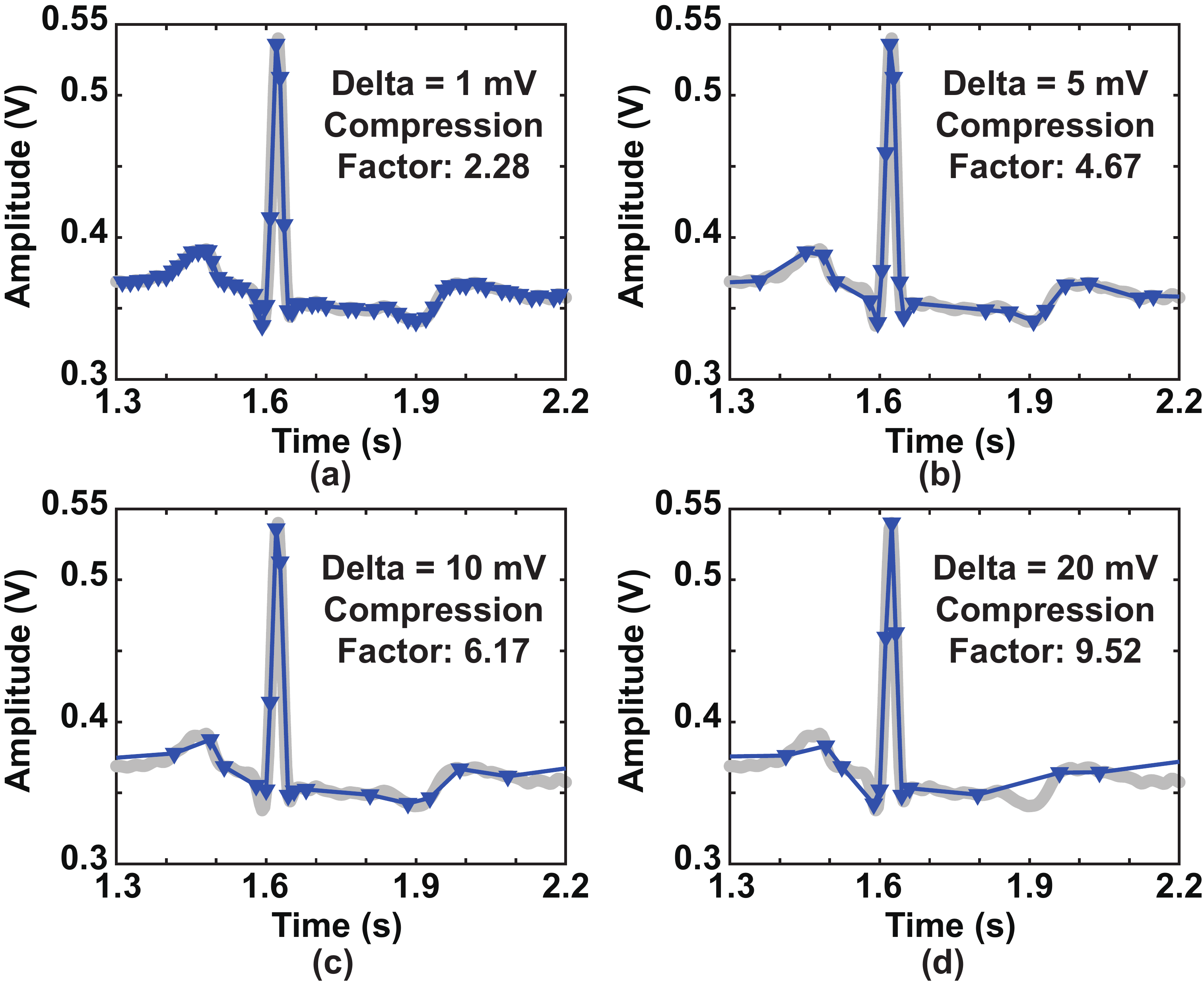}}
	\vspace{-0.1in}
    \caption{Reconstructed ECG waveform by the selected sampling points with different Delta values and the results of compression factors. }\label{ecgexample}
    	\vspace{-0.2in}
\end{figure}


The performance of data saving can be evaluated by the compression factor, which is defined as the ratio of the total data amount generated by Nyquist rate sampling to the data amount from the proposed DPS method \cite{HafshejaniElmiMirabbasiTCASI2020}. The compression factor depends on the amplitude of the signal, the signal sparsity, and the Delta value. If the Delta value is too large, the reconstructed signal may be distorted. To evaluate data-saving performance on biomedical signals, ECG data from the MIT-BIH Arrhythmia database is applied as the input signal.  The simulation result in Fig. \ref{ecgexample} shows that for an ECG signal with 0.3Vpp amplitude a 10 mV Delta can achieve a compression factor of 6.17 while the reconstructed waveform is acceptable for arrhythmia classification. A 20 mV Delta with a compression factor of 9.52 is not acceptable since the \textcolor{blue}{reconstructed} signal is distorted.


\begin{figure}[!t]
	\centerline{\includegraphics[width=3.4in]{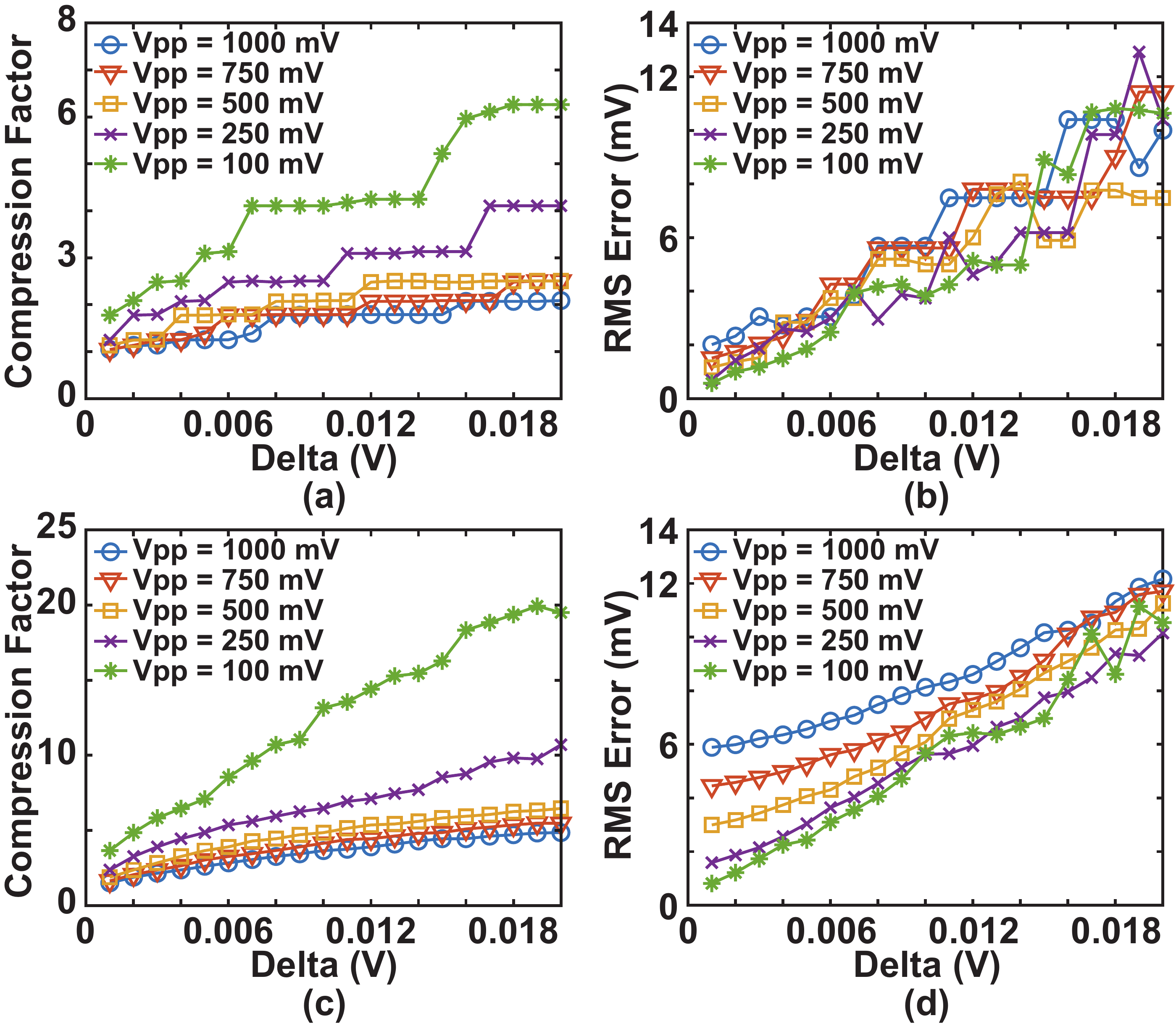}}
	\vspace{-0.1in}
    \caption{Compression Factor and RMS Error at different Delta values for sinusoidal signal (a) (b) and ECG signal (c) and (d) with different amplitudes.}\label{errorDataSaving}
    	\vspace{-0.2in}
\end{figure}

The reconstructed signal from the Dynamic Predictive Sampling method may introduce extra error since it contains fewer sampling points than Nyquist sampling. Reconstruction of the analog signal can be achieved using linear  or polynomial interpolation. In this study, we assume the reconstruction is done by the first-order piece-wise-linear method by simply connecting the selected sampling points using straight lines. Both ECG signals and sinusoidal signals are studied as inputs to evaluate the performance of the proposed DPS method. Fig. \ref{errorDataSaving} (a) and (b) show data saving and root mean square (RMS) error as a function of Delta values at different signal amplitudes for the sinusoidal signal, while Fig. \ref{errorDataSaving} (c) and (d) illustrate performance for the ECG signal. Since the DPS system requires extra timestamp output, we assume the data output of each sample is 10-bits and the timestamp between two samplings is also 10-bits. Thus, each sampling from DPS needs 20-bits while each sampling from Nyquist rate ADC needs only 10-bits. Thanks to the signal sparsity, the proposed DPS method can achieve a high compression factor while keeping the RMS error acceptable.


\vspace{-0.2in}
\subsection{Power Savings}

\begin{figure}[!t]
	\centerline{\includegraphics[width=2.8in]{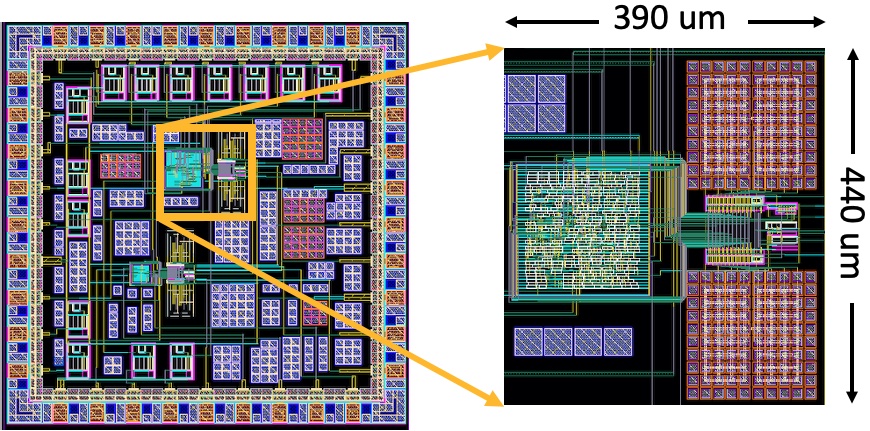}}
	\vspace{-0.1in}
    \caption{Layout of the DPS ADC chip and the core area of the circuits.}\label{layout}
    \vspace{-0.3in}
\end{figure}

To better evaluate the power savings of the proposed DPS method, an integrated ADC circuit is designed using 0.18$\mu$m CMOS Process. The circuit is based on a fully-differential SAR ADC with the updated DPS control logic as shown in Fig. \ref{fig:system}. The whole circuit consists of the DPS digital logic, the switched-capacitor array as the DAC, and the counter for counting the time stamp. The circuit is designed using a 1.8 V power supply and the sampling rate is set at 1 kHz targeting monitoring ECG signals. The internal clock is \textcolor{blue}{16} kHz for the DPS and SAR logic. The overall chip is 1.5mm by 1.5mm including the pad frame. The layout of the integrated circuit is shown in Fig. \ref{layout}. The core circuits occupy a 0.44 mm x 0.39 mm area. The circuit also contains a test structure for simulating a conventional SAR ADC to compare power savings.

\begin{figure}[!t]
	\centerline{\includegraphics[width=3.6in]{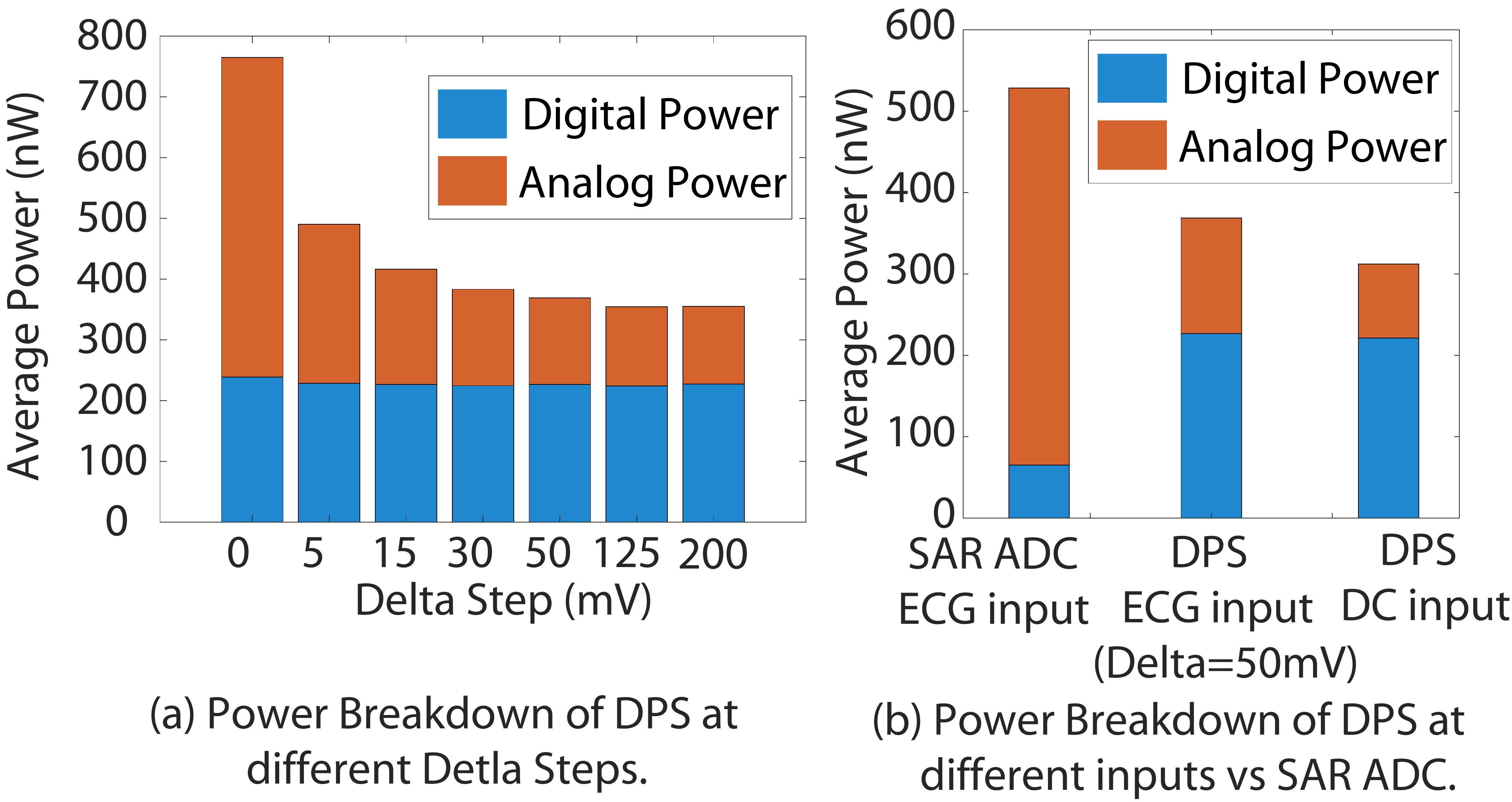}}
	\vspace{-0.1in}
    \caption{(a) power breakdown of the DPS ADC at different Delta value. (b) Comparing power breakdown between the SAR ADC and DPS ADC with the same ECG input, and the DPS ADC with a DC input.}\label{power}
    	\vspace{-0.3in}
\end{figure}

The DPS ADC achieves power saving when the input signal is sparse. The power cost of the DPS ADC depends on both the input signal sparsity and the Delta value. Fig. \ref{power} (a) shows the simulated power breakdown of the DPS ADC at different Delta levels. While the digital power remains constant, the analog power can be saved with a larger Delta value. This is because with a larger Delta value, more predictions are successful. So that the analog power can be greatly saved due to the reduced frequency of performing quantization, which draws power from the comparator. Compared to a conventional SAR ADC, the DPS ADC pays extra effort in selecting sampling points by adding two more comparisons (comparing the input signal with the upper and lower thresholds) and extra calculation of the prediction in the control logic. Power saving is achieved when the prediction is successful so that no further quantization and comparison are required. Therefore, when the Delta value is very small, the DPS ADC may consume more power than a conventional SAR ADC due to the extra comparison and digital operation as shown in Fig. \ref{power} (b). Although a DPS ADC has a larger digital power compared to a SAR ADC, it could be reduced by using advanced fabrication technologies. Since a conventional SAR ADC performs quantization at each sampling, its analog power from the comparator makes its total power higher than a DPS ADC when the input signal is sparse. A power-saving factor can be calculated by comparing the DPS ADC over a conventional SAR ADC at the same sampling rate and input signal. 






\vspace{-0.05in}
\section{Discussion}\label{sec:discussion}

The proposed DPS ADC provides a nonuniform sampling method to reduce the number of quantization in data acquisition systems. It selects only the key sampling points in the input analog waveform for quantization to reduce data throughput and power consumption of the digital processing circuits in the system. The proposed dynamic predictive sampling provides unique features compared to other nonuniform sampling systems such as level-crossing samplings \cite{HouYousefLianTCASII2018} and slope-based signal-dependent samplings \cite{HafshejaniElmiMirabbasiTCASI2020}. The main difference between dynamic predictive sampling and level-crossing sampling is that in the DPS system, the thresholds are updated after every sampling using two prior sampling or prediction values. Even if the prediction is successful, the threshold values in the DPS system change at every sampling clock cycle and tracks the input analog slope. While in the level-crossing sampling, if a sampling event happens, the next thresholds are set to a prior defined fixed value.


\begin{table}[]
\caption{Nonuniform Sampling ADC Performance Comparison }
\label{tab:compare}
\begin{tabular}{|c|c|c|c|}
\hline
 & This work & TCASII'18 \cite{HouYousefLianTCASII2018} & TCASI'20 \cite{HafshejaniElmiMirabbasiTCASI2020} \\ \hline
Method & \begin{tabular}[c]{@{}c@{}}Dynamic \\ Predictive \\ Sampling\end{tabular} & \begin{tabular}[c]{@{}c@{}}Level-crossing\\ Sampling\end{tabular} & \begin{tabular}[c]{@{}c@{}}Slope-based\\ Signal Dependent\\ Sampling\end{tabular} \\ \hline
\begin{tabular}[c]{@{}c@{}}Need\\ Division\end{tabular} & No & No & Yes \\ \hline
\begin{tabular}[c]{@{}c@{}}Turning points\\ Sampling\end{tabular} & Yes & No & Yes \\ \hline
\begin{tabular}[c]{@{}c@{}}Sampling\\ Value\end{tabular} & Real & \begin{tabular}[c]{@{}c@{}}Only at\\ Fixed Levels\end{tabular} & Real \\ \hline
Technology & 180nm & 180nm & 180nm \\ \hline
\begin{tabular}[c]{@{}c@{}}Sampling\\ Rate\end{tabular} & 1 kHz & 1 kHz & 1 kHz \\ \hline
Power (nW) & 368 & 255 & 1700 \\ \hline
Resolution & 10-bit & 6.2-7.9 & 12 \\ \hline
\begin{tabular}[c]{@{}c@{}}Core\\ Area (mm)\end{tabular} & 0.39x0.44 & 0.06x0.24 & 0.5x0.27 \\ \hline
\begin{tabular}[c]{@{}c@{}}Compression\\ Factor\end{tabular} & 6.17 & N/A & 6.1 \\ \hline
\begin{tabular}[c]{@{}c@{}}Power Saving\\ Factor\end{tabular} & 30.3\% & N/A & up to 81\% \\ \hline
\end{tabular}
\vspace{-0.25in}
\end{table}

A comparison between the dynamic predictive sampling method, the level-crossing sampling method, and the  slope-based signal-dependent sampling method is summarized in Table \ref{tab:compare}. Compared to the level-crossing sampling \cite{HouYousefLianTCASII2018}, the DPS method consumes more power. However, level-crossing sampling can only record limited signal amplitude values while DPS reports accurate digital values for each selected sampling point. DPS is more efficient and accurate in terms of localizing turning points in the analog input waveform. More importantly, when considering the reconstruction of the input signal, the DPS method avoids the shifting error due to the insertion and deletion of pulses from the level-crossing sampling system. Furthermore, the level-crossing sampling system often requires the comparator to run in an ``always-on" mode, while the DPS system can turn off the comparator to save power thanks to synchronous operation.

Both DPS and slope-based signal-dependent sampling \cite{HafshejaniElmiMirabbasiTCASI2020} achieves data saving for sparse signal and both are able to record accurate value of turning points. The difference is that the DPS method does not need to perform division in analog circuitry since all calculations are done digitally in DPS. In particular, the DPS system achieves a low computing overhead since multiplying by 2 can be realized easily using shift registers. Moreover, the DPS can be implemented by simply modifying the digital logic in a conventional SAR ADC. In contrast, the slope-based signal-dependent sampling method requires complicated analog circuitry, which is prone to noise and other error effects. From power estimation, the slope-based signal-depended sampling consumes more power than the DPS method due to its complicated analog operation. Furthermore, in the slope-based signal-depended sampling, an accumulated error could occur if the input signal changes slope slowly, while the DPS method provides a guaranteed maximum error, which is the Delta value.

\vspace{-0.1in}
\section{Conclusion}

This paper presents a dynamic predictive sampling method that selects key sampling points in the analog waveform for quantization. The proposed method utilize the sparsity of the input signal to achieve data saving and power saving. An integrated circuits was designed to simulate and analyze power consumption using 0.18 $\mu$m CMOS process. The results show that the proposed system achieves a compression factor of 6.17 and a power saving factor of 30\% for ECG signal. The performance was compared to the conventional SAR ADC, the level-crossing sampling ADC, and the slope-based signal-depended sampling ADC. Compared to the level-crossing sampling ADC, the proposed system has the advantage of recording accurate digital value of key sampling points. It avoids complicated slope calculation circuitry in the slope-based signal-depended sampling ADC. The proposed system can be implemented by modifying digital logic of a conventional SAR ADC while greatly reduces data throughput and power consumption in data acquisition and processing system.

\bibliographystyle{IEEEtran}
\bibliography{allbib,Tang_Publication}

\end{document}